\documentclass[twocolumn,showpacs,preprintnumbers,amsmath,amssymb]{revtex4}

\usepackage{graphicx}
\usepackage{bm}
\usepackage{color}
\usepackage{amsmath}
\usepackage{amsfonts}
\usepackage{amssymb}
\usepackage{epstopdf}
\usepackage{dcolumn}
\usepackage{here}
\usepackage{ulem}

\begin{document}

\preprint{APS/123-QED}

\title{Valence-Bond-Glass State with Singlet Gap in the Spin\,-\,1/2 Square-Lattice\\Random $\bm {J_1\,{-}\,J_2}$ Heisenberg Antiferromagnet Sr$_2$CuTe$_{1-x}$W$_x$O$_6$}

\author{Masari Watanabe$^1$}
\email{watanabe.m.bm@m.titech.ac.jp}
\author{Nobuyuki Kurita$^1$} 
\author{Hidekazu Tanaka$^1$}
\email{tanaka@lee.phys.titech.ac.jp}
\author{Wataru Ueno$^2$}
\author{Kazuki Matsui$^2$}
\author{Takayuki Goto$^2$}

\affiliation{
$^1$Department of Physics, Tokyo Institute of Technology, Meguro-ku, Tokyo 152-8551, Japan\\
$^2$Physics Division, Sophia University, Chiyoda-ku, Tokyo 102-8544, Japan
}
\date{\today}

\begin{abstract}

The double-perovskite compounds Sr$_2$CuTeO$_6$ and Sr$_2$CuWO$_6$ are magnetically described as quasi-two-dimensional spin-1/2 square-lattice $J_1{-}J_2$ Heisenberg antiferromagnets with predominant $J_1$ and $J_2$ exchange interactions, respectively. We report the low-temperature magnetic properties of Sr$_2$CuTe$_{1-x}$W$_x$O$_6$ with randomness in the magnitudes of $J_1$ and $J_2$. It was found that the low-temperature specific heat for $0.1\,{\leq}\,x\,{\leq}\,0.5$ has a large component proportional to the temperature $T$ above 1.2 K, although the low-temperature specific heat for the two parent systems is approximately proportional to $T^3$. With decreasing temperature below 1.2 K, the $T$-linear component decreases rapidly toward zero, which is insensitive to the magnetic field up to 9 T. This is suggestive of the singlet excitation decoupled from the magnetic field. The NMR spectrum for $x\,{=}\,0.2$ exhibits no long-range order down to 1.8 K. These results indicate that the ground state of Sr$_2$CuTe$_{1-x}$W$_x$O$_6$ is a valence-bond-glass state with singlet gaps.

\end{abstract}


\maketitle

\section{Introduction}
In most magnets except for one-dimensional (1D) magnets, the ordered ground state is robust and can survive even in a spin-1/2 triangular-lattice Heisenberg antiferromagnet \cite{Huse,Jolicoeur,Bernu,Singh1,White}, which is a prototypical frustrated quantum magnet. Thus, the disordered ground state induced by the quantum fluctuation has been of great interest and is one of the central topics in condensed matter physics. Quantum disordered ground states (QDGSs) such as the spin liquid state \cite{Kitaev,Balents} and valence-bond-solid state \cite{Read1} have been predicted to exist in frustrated quantum magnets such as spin-1/2 kagome-lattice Heisenberg antiferromagnets \cite{Wang,Hermele,Yan,Depenbrock,Nishimoto,Singh2,Evenbly,Hwang1}.

Recently, it has been theoretically demonstrated that the exchange randomness in frustrated quantum magnets suppresses the spin ordering and induces a QDGS \cite{Watanabe,Kawamura,Shimokawa,Uematsu}. This randomness-induced QDGS is considered to be composed of randomly frozen singlets, which are formed between not only nearest-neighbor spins but also distant spins. This QDGS is termed the random singlet \cite{Dasgupta,Bhatt,Fisher,Lin} or valence-bond-glass (VBG) state \cite{Tarzia_EPL2008,Singh_PRL2010}. The randomness-induced QDGS is characterized by a finite magnetic susceptibility and a low-temperature specific heat proportional to the temperature $T$, which arise from the many singlet spin pairs that can be easily excited to triplets with a small or zero energy \cite{Watanabe,Kawamura,Shimokawa}. The suppression of spin ordering and the $T$-linear specific heat caused by exchange randomness were observed in the spin-1/2 spatially anisotropic triangular-lattice antiferromagnet Cs$_2$CuBr$_{4-x}$Cl$_x$~\cite{Ono_JPSJ2005}, kagome-lattice antiferromagnet (Rb$_{1-x}$Cs$_{x}$)$_2$Cu$_3$SnF$_{12}$~\cite{Katayama} and a honeycomb-lattice organic magnet with random competing exchange interactions~\cite{Yamaguchi}. However, the systematic changes in the ground states and low-temperature magnetic properties upon varying the exchange randomness have not been sufficiently elucidated. 

The spin-1/2 square-lattice Heisenberg antiferromagnet with the nearest-neighbor ($J_1$) and next-nearest-neighbor ($J_2$) exchange interactions, referred to as the $S\,{=}\,1/2$ $J_1{-}J_2$ SLHAF, is a prototypical quantum magnet with bond frustration. The most noteworthy point of this model is that a QDGS emerges in the range of $\alpha_{c1}\,{<}\,J_2/J_1\,{<}\,\alpha_{c2}$ with $\alpha_{c1}\,{\simeq}\,0.4$ and $\alpha_{c2}\,{\simeq}\,0.6$ \cite{Chandra,Dagotto,Figueirido,Read2,Igarashi,Einarsson,Zhitomirsky,Bishop,Sirker,Mambrini,Darradi}. The ground states for $J_2/J_1\,{<}\,\alpha_{c1}$ and $\alpha_{c2}\,{<}\,J_2/J_1$ are N\'{e}el antiferromagnetic and collinear antiferromagnetic states, respectively. In this paper, we report the QDGS observed in the spin-1/2 square-lattice random $J_1{-}J_2$ Heisenberg antiferromagnet Sr$_2$CuTe$_{1-x}$W$_x$O$_6$ with $0.1\,{\leq}\,x\,{\leq}\,0.5$.  

\begin{figure}[t!]
\begin{center}
\includegraphics[width=0.95\linewidth]{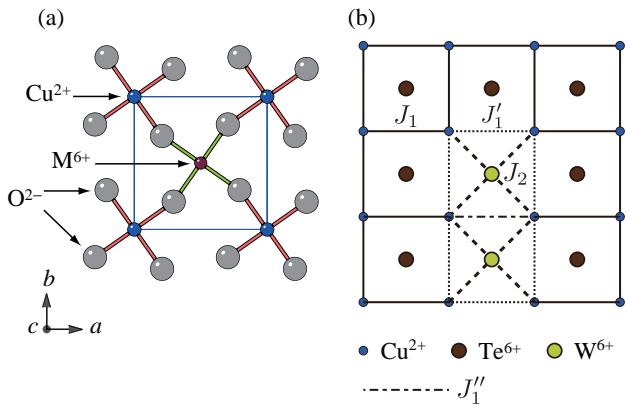}
\end{center}
\vspace{-15pt}\caption{(Color online) (a) Linkage of CuO$_6$ and MO$_6$ octahedra in the $ab$ plane of Sr$_2$CuMO$_6$ (M$\,{=}\,$W and Te). (b) Illustration of exchange interactions around W$^{6+}$ ions that substitute for  Te$^{6+}$ ions.}
\label{structure}
\end{figure}

The two parent compounds, Sr$_2$CuWO$_6$ and Sr$_2$CuTeO$_6$, have the tetragonal structure, in which CuO$_6$ and MO$_6$ octahedra are arranged alternately in the $ab$ plane, sharing their corners as shown in Fig.~\ref{structure}(a). Because the hole orbitals $d(x^2-y^2)$ of Cu$^{2+}$ ions with spin-1/2 are spread in the $ab$ plane, exchange interactions in the $ab$ plane are much stronger than those between the $ab$ planes. Consequently, Sr$_2$CuWO$_6$ and Sr$_2$CuTeO$_6$ are described as quasi-2D $S\,{=}\,1/2$ $J_1{-}J_2$ SLHAFs \cite{Iwanaga,Vasala3,Koga}.

Sr$_2$CuWO$_6$ and Sr$_2$CuTeO$_6$ undergo three-dimensional magnetic orderings at $T_{\rm N}\,{=}\,24$ and 29 K, respectively, owing to the weak interlayer exchange interactions \cite{Vasala3,Vasala2,Koga2}. However, the spin structures in their ordered states are different. The collinear antiferromagnetic and N\'{e}el antiferromagnetic states are realized in Sr$_2$CuWO$_6$ \cite{Vasala2} and Sr$_2$CuTeO$_6$ \cite{Koga2}, respectively. This indicates that the dominant exchange interaction is $J_2$ in Sr$_2$CuWO$_6$, while $J_1$ is dominant in Sr$_2$CuTeO$_6$. This difference can be understood from the difference in the electronic states of the outermost filled orbital of the nonmagnetic M$^{6+}$ \cite{Koga2,Yokota}. Thus, we expect that the partial substitution of W$^{6+}$ for Te$^{6+}$ will produce the randomness in $J_1$ and $J_2$ interactions, which leads to the VBG state. With this motivation, we synthesized Sr$_2$CuTe$_{1-x}$W$_x$O$_6$ samples with various tungsten concentrations $x$ and investigated their low-temperature magnetic properties.

\vspace{-5pt}
\section{Experimental details}

Powder samples of Sr$_2$CuTe$_{1-x}$W$_x$O$_6$ were synthesized from mixtures of SrCO$_3$, CuO, TeO$_2$, and WO$_3$ with molar ratios of $2\,{:}\,1\,{:}\,1\,{-}\,x\,{:}\,x$ by a solid-state reaction. Each mixed powder was ground well with an agate mortar and fired at 1000\,$^{\circ}$C in air for 24\,h. The powder was then reground, pelletized, and calcined twice at 1000\,$-$\,1100\,$^{\circ}$C for 24\,h in an oxygen atmosphere. Using X-ray powder diffraction, we confirmed that W$^{6+}$ ions substitute for Te$^{6+}$ ions in Sr$_2$CuTe$_{1-x}$W$_x$O$_6$ and that the phase separation of the two parent compounds does not occur. 

The magnetic susceptibilities of the Sr$_2$CuTe$_{1-x}$W$_x$O$_6$ powders were measured down to 1.8 K using a SQUID magnetometer (Quantum Design MPMS XL). The specific heat of the Sr$_2$CuTe$_{1-x}$W$_x$O$_6$ powders was measured down to 0.36 K in magnetic fields of up to 9 T using a physical property measurement system (Quantum Design PPMS) by the relaxation method.
Nuclear magnetic resonance (NMR) measurements were performed on a powder sample with $x\,{=}\,0.2$ using a 16 T superconducting magnet in the temperature range between 1.8 and 20 K.

\vspace{-5pt}
\section{Results and Discussion}
\begin{figure}[t]
\begin{center}
\includegraphics[width=1.0\linewidth]{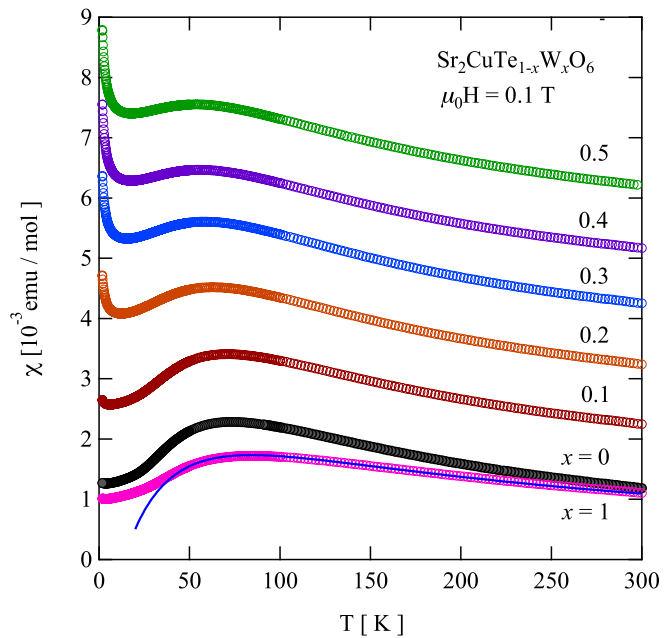}
\end{center}
\vspace{-15pt}\caption{(Color online) Temperature dependence of magnetic susceptibility of Sr$_{2}$CuTe$_{1-x}$W$_{x}$O$_{6}$ measured at $\mu_{0}H\,{=}\,0.1$ T for various $x$. The susceptibility data for $0.1\,{\leq}\,x\,{\leq}\,0.5$ are shifted upward by multiples of $1\,{\times}\,10^{-3}$ emu/mol. The solid line superimposed on the data for $x\,{=}\,1$ shows the susceptibility calculated by the Pad\'{e} approximation with $J_1/k_{\rm B}\,{=}\,22.6$ K, $J_2/k_{\rm B}\,{=}\,91.2$ K, and $g\,{=}\,2.18$.}
\label{sus}
\end{figure}

Figure \ref{sus} shows the temperature variation of the magnetic susceptibility for $0\,{\leq}\,x\,{\leq}\,0.5$ and $x\,{=}\,1$ measured at $\mu_{0}H\,{=}\,0.1$\ T. The susceptibility data of the two parent compounds ($x\,{=}\,0$ and 1) coincide with those reported in Refs. \cite{Koga} and \cite{Vasala3}, respectively. With decreasing temperature, the susceptibilities of the parent compounds display broad maxima at approximately $T_{\rm max}\,{=}\,70-90$ K owing to the short-range spin correlation. This susceptibility behavior is characteristic of two-dimensional $S\,{=}\,1/2$ SLHAFs\,\cite{de-Jongh,Kim,Rosner}. With further decreasing temperature, the magnetic susceptibilities for $x\,{\neq}\,0$ and 1 exhibit a Curie-like upturn. As the tungsten concentration $x$ increases, the upturn is more enhanced, which gives rise to a shift of $T_{\rm max}$ toward the low-temperature side. This upturn probably originates from almost free or uncoupled spins, which are produced by exchange randomness. We also measured field-cooled (FC) and zero-field-cooled (ZFC) magnetic susceptibilities to clarify whether or not the ground state is the ordinary spin-glass state. Because no significant difference was found between the FC and ZFC data, the possibility of a spin-glass ground state is ruled out.

\begin{figure}[h]
\begin{center}
\includegraphics[width=1.0\linewidth]{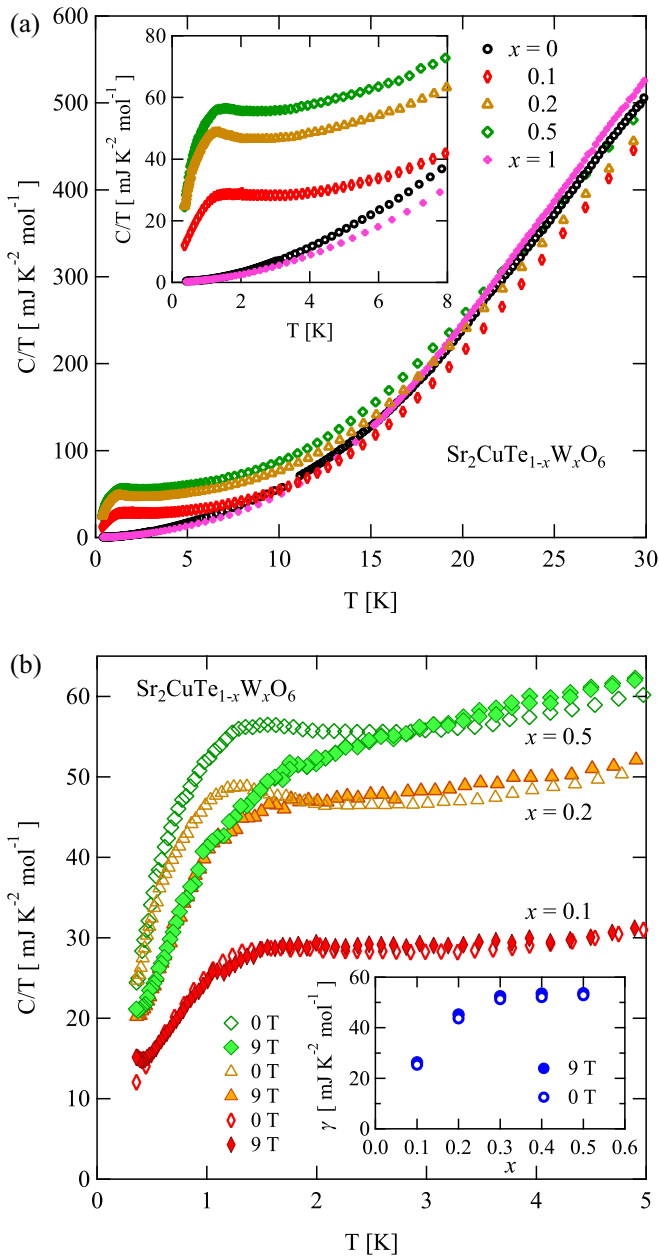}
\end{center}
\vspace{-15pt}\caption{(Color online) (a) Temperature dependence of $C/T$ of Sr$_{2}$CuTe$_{1-x}$W$_{x}$O$_{6}$ powders measured at zero magnetic field for $x\,{=}\,0, 0.1, 0.2, 0.5$, and 1. The inset shows an enlargement of the data below 8 K. 
(b) Temperature dependence of $C/T$ for $x\,{=}\,0.1, 0.2$, and 0.5 measured at zero field (open symbols) and 9~T (closed symbols). The inset shows the coefficient $\gamma$ of the $T$-linear component vs $x$ obtained at zero field and 9~T.}
\label{heat}
\end{figure}

The nearest-neighbor interaction $J_1$ in Sr$_2$CuTeO$_6$ was evaluated as $J_1/k_{\rm B}\,{=}\,80$ and 83 K from the magnetic susceptibility by the Pad\'{e} approximation \cite{Koga} and from the dispersion relations of magnetic excitations \cite{Babkevich}, respectively. The next-nearest-neighbor interaction $J_2$ is negligible in Sr$_2$CuTeO$_6$. However, for Sr$_2$CuWO$_6$, the susceptibility data have only been analyzed by the classical molecular-field approximation \cite{Iwanaga}. Here, we estimate the exchange constants of Sr$_2$CuWO$_6$ from the susceptibility data by the $[5, 5]$ Pad\'{e} approximation using the result of the high-temperature expansion of ${\beta}\,{=}\,1/k_{\rm B}T$ up to the tenth order \cite{Rosner}. We assume that $J_2\,{>}\,J_1$ in Sr$_2$CuWO$_6$. The best fit between 45 and 300 K is obtained with $J_1/k_{\rm B}\,{=}\,22.6$ K and $J_2/k_{\rm B}\,{=}\,91.2$ K using $g\,{=}\,2.18$, which was determined from the paramagnetic resonance. The solid line in Fig.~\ref{sus} shows the susceptibility calculated with these parameters. 

\begin{figure}[t]
\begin{center}
\includegraphics[width=1.0\linewidth]{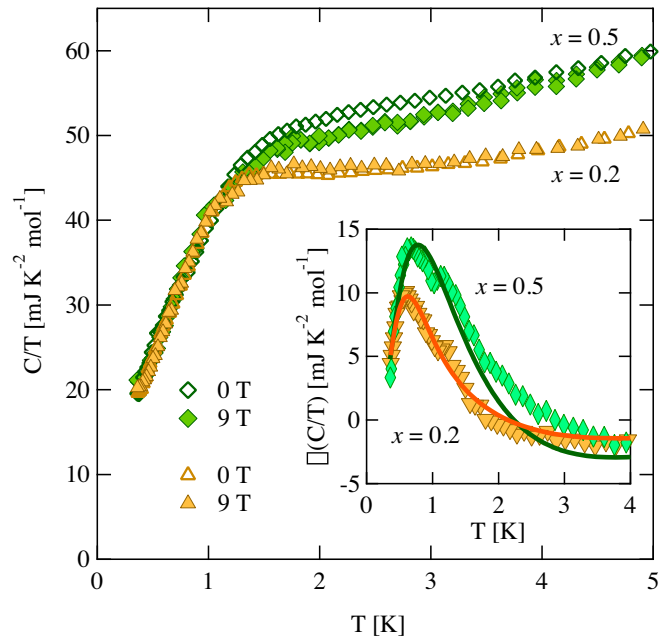}
\end{center}
\vspace{-15pt}\caption{(Color online)  Temperature dependence of $C/T$ for $x\,{=}\,0.2$, and 0.5 measured at zero field and 9~T after the subtraction of the Schottky specific heat due to the Zeeman splitting of loosely coupled spins. The inset shows the temperature dependence of difference ${\Delta}(C/T)$ between $C/T$ measured at $\mu_{0}H\,{=}\,0$ and 9 T for $x\,{=}\,0.2$, and 0.5. Solid lines are the ${\Delta}(C/T)$ for the Schottky specific heat calculated with parameters shown in the text.}
\label{heat_2}
\end{figure}

Figure~\ref{heat}(a) shows the temperature dependence of the specific heat divided by the temperature $C/T$ for $x\,{=}\,0, 0.1, 0.2, 0.5$, and 1 measured at zero magnetic field. No anomaly indicative of magnetic ordering was observed even for the parent compounds Sr$_2$CuTeO$_6$ and Sr$_2$CuWO$_6$, which undergo magnetic phase transitions at $T_{\rm N}\,{=}\,29$~K~\cite{Koga2} and 24~K~\cite{Vasala3,Vasala2}, respectively.
We can see from the inset of Fig.~\ref{heat}(a) that the partial substitution of W$^{6+}$ for Te$^{6+}$ causes a large change in the  low-temperature specific heat. The low-temperature specific heat $C(T)$ of the parent compounds ($x\,{=}\,0$ and 1) is described as $C(T)\,{=}\,{\beta}T^{2}\,{+}\,{\alpha}T^{3}$, where the $T^{2}$ component is much smaller than the $T^{3}$ component. The small $T^{2}$ component can be attributed to quasi-2D magnetic excitations of the parent compounds. In contrast to the specific heat of the parent compounds, a $T$-linear component is clearly observed in the low-temperature specific heat for $0.1\,{\leq}\,x\,{\leq}\,0.5$ above 1.5 K. With decreasing temperature from 1.5 K, the $T$-linear component decreases rapidly toward zero, resulting in a shoulder anomaly appearing in $C/T$ near 1.2 K. 

Figure~\ref{heat}(b) shows $C/T$ for $x\,{=}\,0.1, 0.2$, and 0.5 measured at $\mu_{0}H\,{=}\,0$ and 9 T. The shoulder anomaly in $C/T$ observed at zero magnetic field is partly suppressed at 9 T for $x\,{=}\,0.2$ and 0.5, whereas for $x\,{=}\,0.1$, $C/T$ is almost independent of the applied magnetic field. 
Above 2 K, no significant difference is observed in the specific heat data measured at $\mu_{0}H\,{=}\,0$ and 9 T. Applying the formula $C(T)\,{=}\,{\gamma}T\,{+}\,{\alpha}T^{3}$ for $3\,{\leq}\,T\,{\leq}\,7$ K, we estimate the coefficient $\gamma$ of the $T$-linear component. The inset of Fig.~\ref{heat}(b) shows $\gamma$ as a function of $x$. $\gamma$ increases with increasing $x$ and saturates at ${\gamma}\,{\simeq}\,54$ mJ/(K$^2$\,mol) around $x\,{=}\,0.3$. Recent theory~\cite{Watanabe,Kawamura,Shimokawa,Uematsu} has demonstrated that the specific heat has a $T$-linear component in frustrated quantum magnets with random bonds, which arises from the low-energy gapless excitations. The maximum $\gamma$ value in Sr$_2$CuTe$_{1-x}$W$_x$O$_6$ is the same order of magnitude as that calculated for the $S\,{=}\,1/2$ triangular-lattice random bond Heisenberg antiferromagnet~\cite{Watanabe}. 

 From the analysis shown below, we deduce that the difference between the values of $C/T$ at $\mu_{0}H\,{=}\,0$ and 9 T below 2 K arises from the Schottky specific heat due to the Zeeman splitting of loosely coupled spin pairs. 
The inset of Fig.~\ref{heat_2} shows the temperature dependence of the difference ${\Delta}(C/T)$ between $C/T$ measured at $\mu_{0}H\,{=}\,0$ and 9 T for $x\,{=}\,0.2$, and 0.5. With increasing temperature from 0.36 K, ${\Delta}(C/T)$ displays a rounded maximum at $0.6\,{-}\,0.7$ K and decreases to be negative. This is typical of the Schottky specific heat due to the Zeeman splitting. Specific heat of spin pairs coupled via exchange interaction $J$ in magnetic field is expressed as 
\begin{widetext}
\begin{eqnarray}
&&C(T,H,J) =\nonumber\\
&&\frac{nN_{\rm A}{\beta}}{2T}\,\frac{2\hspace{-0.5mm}\left\{\left(J^2+(g\mu_{\rm B}H)^2\right)e^{{\beta}J}+(g\mu_{\rm B}H)^2\right\}\cosh\left({\beta}{g\mu_{\rm B}H}\right)-4Jg\mu_{\rm B}H e^{{\beta}J} \sinh\left({\beta}{g\mu_{\rm B}H}\right)+4(g\mu_{\rm B}H)^2+J^2e^{{\beta}J}}{\left(1+e^{{\beta}J}+2\cosh({\beta}{g\mu_{\rm B}H})\right)^2},\hspace{10mm}
\label{C_dimer}
\end{eqnarray}
\end{widetext}
where $N_{\rm A}$ is the Avogadro's number and $n$ is the fraction of the spins that are loosely coupled to form spin pairs. In this analysis, we assume that the coupling constant $J$ is uniformly distributed between $J\,{-}\,{\Delta}J$ and $J\,{+}\,{\Delta}J$. Solid lines in the inset of Fig.~\ref{heat_2} are ${\Delta}(C/T)\,{=}\,C(0\,{\rm T})/T\,{-}\,C(9\,{\rm T})/T$  for $x\,{=}\,0.2$, and 0.5 calculated with $n\,{=}\,1.7\,{\times}\,10^{-3}$, $J/k_{\rm B}\,{=}\,2.3$~K and ${\Delta}J/k_{\rm B}\,{=}\,0.63$~K, and $n\,{=}\,3.4\,{\times}\,10^{-3}$, $J/k_{\rm B}\,{=}\,3.1$\,K and ${\Delta}J/k_{\rm B}\,{=}\,1.3$\,K, respectively. The $g$ factor is set to be $g=2.18$. Experimental results of ${\Delta}(C/T)$ are well described in terms of the Schottky specific heat due to the Zeeman splitting of loosely coupled spin pairs. Because the coupling constant $J$ is of the order of 1 K, these spins give rise to the Curie-like term in the magnetic susceptibility at low temperatures. The fraction of spins that produce the Curie-like term is estimated as $n\,{=}\,1.9\,{\times}\,10^{-3}$ and $3.5\,{\times}\,10^{-3}$ for $x\,{=}\,0.2$, and 0.5, respectively, which coincide with those obtained from the analysis of low-temperature specific heat. Thus, we can deduce that the magnetic field dependence of the low-temperature specific heat and the Curie-like term in the magnetic susceptibility arise from the loosely coupled spin pairs.

 Figure~\ref{heat_2} shows $C/T$ at $\mu_{0}H\,{=}\,0$ and 9 T for $x\,{=}\,0.2$, and 0.5 after the correction of the Schottky specific heat. Down to the lowest temperature of 0.36 K, no significant difference is observed in $C/T$ at $\mu_{0}H\,{=}\,0$ and 9 T.
The shoulder anomaly around 1.2 K is not sharp and the temperature that gives the shoulder anomaly is almost independent of the tungsten concentration $x$. The shoulder anomaly and the rapid decrease in $C/T$ toward zero persist even at 9 T, the Zeeman energy of which is much larger than the energy corresponding to 1.2 K. These results indicate that the shoulder anomaly in $C/T$ cannot be ascribed to the magnetic ordering or spin-glass transition. Because the low-temperature specific heat is insensitive to the magnetic field, we deduce that the shoulder anomaly originates from the singlet excitations, which are decoupled from the magnetic field. 

\begin{figure}[t]
\begin{center}
\includegraphics[width=0.95\linewidth]{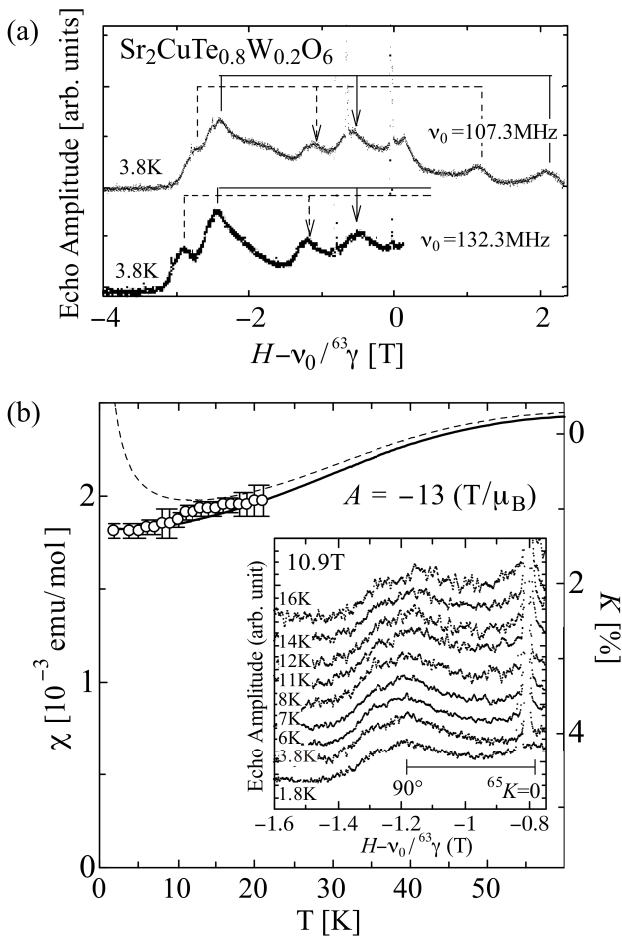}
\end{center}
\vspace{-15pt}\caption{(a) $^{63/65}$Cu-NMR spectra measured in magnetic field range between 6 and 12 T. The abscissa is shifted by $H\,{-}\,{\nu}_0/\,^{63}{\gamma}$, where ${\nu}_0$ and $\gamma$ are the NMR frequency and the nuclear gyromagnetic ratio, respectively. The vertical solid and dashed lines (arrows) denote the satellite (central) transition positions for $^{63}$Cu and $^{65}$Cu nuclei, respectively.
(b) Temperature dependence of the Knight shift (open symbols) scaled with the uniform susceptibility, from which the Curie term is subtracted (solid curve).  The raw data for the uniform susceptibility are shown by the dashed curve. The inset shows typical spectral profiles at various temperatures. The horizontal line denotes the position of the $90^{\circ}$ peak for the $^{65}$Cu central transition and the zero-shift position.}
\label{NMR}
\end{figure}

Figure \ref{NMR}(a) shows two typical $^{63/65}$Cu-NMR spectra measured with two different frequencies ${\nu}_0$. One can observe well-resolved peaks corresponding to the singular points in the quadrupolar powder pattern for the $I\,{=}\,3/2$ nuclei. To extract $K$, which is the in-plane component of the Knight shift, from the observed spectra, we measured the positions of the 90$^{\circ}$ peak for the $^{65}$Cu central transition denoted by the dashed arrows in Fig.~\ref{NMR}(a) at ten different frequencies between 109 and 132 MHz and analyzed them using the second-order perturbation formula~\cite{Inoue}. The values of $K$ and the nuclear quadrupolar parameter $^{63}{\nu}_{\rm Q}$ at 3.8 K were determined to be 1.3\% and 52 MHz, respectively. The positive value of $K$ suggests that it includes an appreciable orbital contribution.

Next, to obtain the temperature dependence of $K$, we traced the peak position.  Typical spectral profiles are shown in the inset of Fig.~\ref{NMR}(b). No anomalous broadening or splitting, indicative of a magnetic order, was observed down to 1.8 K. Assuming that $^{63}{\nu}_{\rm Q}$ is temperature-independent at low temperatures, the temperature dependence of $K$ was simply determined from the peak position and is shown in Fig.~\ref{NMR}. It decreased with decreasing temperature and scales well with the uniform susceptibility, from which the tiny Curie-like term was subtracted. From the scaling factor, the hyperfine coupling constant $A$ was estimated to be $-13$~(T/${\mu}_{\rm B}$). Its negative sign and its magnitude are typical for divalent copper, indicating that the present NMR effectively probes the magnetism of the copper spin. Consequently, we can safely conclude at this stage that the bulk spin susceptibility of this system decreases with decreasing temperature and approaches a finite value at low temperatures.

As shown above, the NMR spectrum for $x\,{=}\,0.2$ indicates the absence of magnetic ordering down to 1.8~K. The small amount of substitution of W$^{6+}$ for Te$^{6+}$ induces marked suppression of the magnetic ordering. The partial substitution of W$^{6+}$ for Te$^{6+}$ also produces a $T$-linear component in the low-temperature specific heat and a Curie term in the magnetic susceptibility. The $T$-linear component and Curie term increase with increasing tungsten concentration $x$.  These properties are characteristic of the VBG state for frustrated quantum magnets with random bonds ~\cite{Watanabe,Kawamura,Shimokawa,Uematsu}. In the VBG state, there are many loosely coupled singlet spins, which can be easily excited. The $T$-linear component in the specific heat and the Curie term in the magnetic susceptibility arise from the low-energy excitations of these spins. Thus, we deduce that the ground state of Sr$_2$CuTe$_{1-x}$W$_x$O$_6$ is the VBG state at least for $0.2\,{\leq}\,x\,{\leq}\,0.5$, as predicted by the theory. It is considered that there is a critical $x_{\rm c}$ that separates the ordered state and the VBG state. In the present experiments, $x_{\rm c}$ was not determined. The data of $C/T$ show a rapid decrease below 1.2~K, which is insensitive to the magnetic field. Thus, we infer that the VBG ground state is accompanied by singlet excitations.

As shown in Fig.~\ref{heat}, the temperature that gives the shoulder in $C/T$ is almost independent of the tungsten concentration $x$. This suggests that the singlet excitations are determined by the local structure of the exchange interactions. Figure~\ref{structure}(b) illustrates the exchange interactions when Te$^{6+}$ ions are substituted by W$^{6+}$ ions. We assume that $J_1$ and $J_2$ are almost the same as those in Sr$_2$CuTeO$_6$ and Sr$_2$CuWO$_6$, respectively, which are $J_1/k_{\rm B}\,{\simeq}\,80$ and $J_2/k_{\rm B}\,{\simeq}\,90$~K. $J_1^{\prime}$ and $J_1^{\prime\prime}$ are the nearest-neighbor exchange interactions via TeO$_6$ and WO$_6$ octahedra and two WO$_6$ octahedra, respectively, which are estimated to be $J_1^{\prime}/k_{\rm B}\,{\simeq}\,50$ and $J_1^{\prime\prime}/k_{\rm B}\,{\simeq}\,20$~K using the exchange constants for Sr$_2$CuWO$_6$ obtained in this study and those for Sr$_2$CuTeO$_6$~\cite{Koga,Babkevich}.

Very recently, Mustonen {\it et al.}~\cite{Mustonen} reported the magnetic properties of Sr$_{2}$CuTe$_{0.5}$W$_{0.5}$O$_{6}$. Their magnetic susceptibility and specific heat data measured down to 2~K are consistent with our data for $x\,{=}\,0.5$. From muon spin relaxation and rotation measurements, they observed the absence of magnetic ordering or a spin-glass transition in Sr$_{2}$CuTe$_{0.5}$W$_{0.5}$O$_{6}$ down to 19~mK. 
\vspace{-15pt}
\section{Conclusion}
In conclusion, we have reported the results of magnetization, specific heat, and NMR measurements on Sr$_2$CuTe$_{1-x}$W$_x$O$_6$, which is characterized as an $S\,{=}\,1/2$ square-lattice random $J_{1}\,{-}\,J_{2}$ Heisenberg antiferromagnet. The partial substitution of W$^{6+}$ for Te$^{6+}$ causes a marked change in the ground state and low-temperature thermodynamic properties. The magnetic ordering observed in the parent compounds is strongly suppressed. The ground state, at least for $0.2\,{\leq}\,x\,{\leq}\,0.5$, is concluded to be the VBG state with a singlet excitation gap of about 1~K.

{\it Note added in proof.} At the proof stage, we noticed that Walker {\it et al}. \cite{Walker} estimated the exchange interactions in Sr$_2$CuWO$_6$ as $J_1/k_{\rm B}$ = 14 K and $J_2/k_{\rm B}$ = 110 K from the magnetic excitation data obtained by inelastic neutron scattering on powdered sample.
\vspace{-15pt}
\section*{ACKNOWLEDGMENTS}

This work was supported by Grants-in-Aid for Scientific Research (A) (No.~17H01142) and (C) (No.~16K05414) from Japan Society for the Promotion of Science.

\end{document}